\documentclass{ws-procs961x669}            % default, citations in superscript
\usepackage{bbm}

\def\d{{\rm d}}

	\newcommand{\beq} {\begin{equation}}
	\newcommand{\eeq} {\end{equation}}

	\newcommand{\B}{\mathcal{B}}
	
	\newcommand{\N}{\mathcal{N}}

	\newcommand{\bea}{\begin{align}}
	\newcommand{\eea}{\end{align}}

\usepackage{tikz}
\usetikzlibrary{decorations.pathreplacing,arrows,positioning,angles,calc,external}
	\tikzset{
    		>=stealth',pil/.style={->,thick,shorten <=2pt,shorten >=2pt,}
		}
 \usepackage{pgfplots}
 	\pgfplotsset{compat=1.3,every axis/.append style={font=\small,thin,tick style={ultra thin}}}
  	\usepgfplotslibrary{fillbetween}

\begin{document}
\title{Wave propagation in the anti-deSitter optical metric}

\author{D. Garc\'ia-Pel\'aez$^*$}
\address{Universidad Aut\'onoma Metropolitana Azcapotzalco,\\
Avenida San Pablo Xalpa 180, Azcapotzalco, Reynosa Tamaulipas, 02200 Ciudad de M\'exico, M\'exico
}
\address{Universidad Panamericana,\\ Tecoyotitla 366.
Col. Ex Hacienda Guadalupe Chimalistac, 01050 M\'exico D.F., M\'exico\\
$^*$E-mail: dgarciap@up.edu.mx}

\author{C. S. L\'opez-Monsalvo$^{**}$}
\address{Conacyt-Universidad Aut\'onoma Metropolitana Azcapotzalco,\\
Avenida San Pablo Xalpa 180, Azcapotzalco, Reynosa Tamaulipas, 02200 Ciudad de M\'exico, M\'exico\\
$^{**}$E-mail: cslopezmo@conacyt.mx}

\begin{abstract}
In this work, we use the fact that kinematics of light propagation in a \emph{non-dispersive medium} associated with a bi-metric spacetime is expressed by means of a 1-parameter family of contact transformations. We present a general technique to find such transformations and explore some explicit examples for Minkowski and anti-deSitter spacetimes geometries.
\end{abstract}


\bodymatter

\section{Introduction}

Soon after the advent of general relativity, it was observed that the gravitational field \emph{bends} the path followed by a light ray. This was confirmed by Eddington after noticing that the apparent position of the stars are shifted from their expected position in the sky when observed during a solar eclipse. This \emph{bending} phenomenon is analogous to the deviation of light rays while traveling in a medium whose refractive index changes from one place to the other. In this sense, it was shown that a  class of optical media can be modeled by means of  a metric tensor  encoding its electromagnetic properties (Refs. \citenum{gordon1923lichtfortpflanzung,de1971gravitational,ehlers2012republication}). Furthermore, this analogy has evolved into the very active field of transformation optics, where the techniques and tools of differential geometry -- that have been insightful in gravitational physics -- have found its way into the more applied area of material science (Refs. \citenum{pendry2006controlling,leonhardt2006general,chen2010transformation,LOPEZMONSALVO2020168270}). In addition,  this has also been used in modeling analogue gravitational spacetimes such as  black holes  and cosmological solutions (Refs. \citenum{PhysRevA.102.023528,schuster2019electromagnetic,faccio2013analogue,schuster2018bespoke}). 
%or to approximate curved spaces by $N$-dimensional \emph{simplices} in order to describe light propagation on them \citenum{GeorgantzisGarcia:20}.\\ 

Fermat's principle poses  a well known problem in the calculus of variations.  In the non-relativistic setting -- where the notion of time is \emph{absolute} and \emph{universal} -- it states that light travels between two given \emph{spatial} locations following a path minimizing a time functional. Such a perspective is clearly untenable in the context of General Relativity, where  it is commonly replaced by the assumption that the path taken by light in traveling from one location to another corresponds to a null geodesic. However, albeit it remains a variational problem, its precise formulation is far more elaborate (see Theorem 7.3.1 in  Ref. \citenum{perlick2000ray}).

Similarly (cf. Theorem 7.1.2 in Ref.\citenum{perlick2000ray}), Huygens' principle is centered in  the idea of light emissions being instantaneous for different observers at each instant in time  (Ref. \citenum{berest1994huygens}). In this way,  light emission is described by well localized wavefronts, as they travel through  three dimensional space. In the relativistic setting,  the electromagnetic field on a precise event on spacetime should depend only on initial conditions originated its past null cone  (Ref. \citenum{harte2013tails}). This has led to explore different wave phenomena where Huygens' principle is not fulfilled, most of which are due to dissipation processes where the wave distribution has not converging tails. As a consequence, the wavefronts cannot be located in exactly at a point of spacetime. With the Hadamard conjecture, this phenomenon has been related to the dimensions and curvature of spacetime   (Refs. \citenum{mclenaghan1969explicit, noonan1995huygens}).

One of the simplest and of most interesting curved spacetimes for theoretical physicists, is the anti-deSitter spacetime (AdS). This spacetime model has a very deep relationship with hyperbolic geometry, where its boundary at spacelike infinity leads  to have strange peculiar like closed timelike curves and achornal surfaces (Refs. \citenum{gibbons2000anti, moschella2006sitter}). Nevertheless, AdS has gained more interest in recent years, as it has showed its value in topics like the \emph{holographic principle}, the ground state of gauged supergravity theories and its central role in high energy physics in the correspondence with the Conformal Field Theory (AdS/CFT). 
 
In this work, we use the fact that an optical medium can be represented by a Riemannian manifold $(\B,g)$ where $\B$ is considered to be the physical space and $g$ the optical spatial metric. The geodesic flow in its unitary tangent bundle can be represented by a contact transformation acting on its space of contact elements. This fact, allows us to describe the wavefront evolution in an optical medium solely in terms of the contact transformation and to reconstruct the geodesic flow from the Reeb vector field. This provides us with a way to construct wavefronts in optical media without directly solving the wave equation. Particularly, we use this powerful tool to reconstruct wave propagation, along with with its corresponding light rays in a $(2+1)$-dimensional AdS spacetime.

This manuscript is structured as follows. In section \ref{sec.stbton} we give the mathematical formalism to construct the technique. In section \ref{sec.lob} we use this technique to reconstruct wavefronts and ray lights in a $(2+1)$-dimensional anti-deSitter spacetime. We find explicitly the 1-parameter family of contact transformations and solve numerically to reconstruct the wave propagation in optical media with interfaces of different refractive index.

\section{Mathematical background}
\label{sec.stbton}

In this work, an optical medium will be a Riemannian manifold $(\mathcal{B}, g)$ submersed in a bi-metric  Lorentzian spacetime $(M,g_0,\tilde g)$. Where $g_0$ is the spacetime metric, used to low and rise index, the metric $\tilde g$ is the optical metric introduced by Gordon  (Ref. \citenum{gordon1923lichtfortpflanzung}), from which the effective speed of light in a medium can be read off.The manifold $\B$ along with his metric $g$, is called \emph{the material manifold}, The metric $g$ can be decomposed in terms of the material manifold Riemannian metric and an observer's four velocity.

The cotangent bundle of the material manifold carries a natural symplectic structure, endowed by a non-degenerated closed 2-form $\omega$. In this sense, the Liouville's vector filed, which preserves the symplectic 2-form in terms of its Lie derivative, defines a 1-form $\lambda \in T^*(T^*\B)$ which generates the symplectic structure. As defined by Perlick (see Definition 5.1.1 and Proposition 5.1.1 in  Ref. \citenum{perlick2000ray}) the punctured cotangent bundle $PT^*\B$, corresponds to a \emph{ray-optical structure} $\N$. Using the inclusion map from this ray-optical structure to the cotangent bundle, the 1-form $\lambda$ induces a contact 1-form $\eta$, where the pair $(\N, \mathcal{D})$ is a contact manifold, known as the \emph{space of contact elements} or the \emph{contact bundle} of $\B$.
 
Let us use the defintion of a \emph{wave front} centered at the point $b \in \B$ as the hypersurface  (Ref. \citenum{HGeiges-ContTopo})
	\beq
	F_b(t)=\{b_i \in \B \,\vert \gamma(0)=b, \gamma(t)=b_i\} 
	\eeq
where $\gamma$ is a geodesic  on $(\B,g)$. For every point in the wave front, the contact element of $\B$ associated to that point is tangent to the wave front. If we consider a geodesic flow on the unitary tangent bundle of $\B$, there exists a unique contact element, which is perpendicular to. So the Reeb vector filed, associated to the contact 1-form of the ray-optical structure is dual to the geodesic flow in the unitary tangent bundle of $\B$. The Reeb vector field is also, the infinitesimal generator of a family of 1-parameter strict contact transformations $\phi_t: \mathcal{N} \to \mathcal{N}$, which preserves the contact form and thus the contact elements on $\B$, which are tangent to the wave fronts. In this sense, the family of transformations $\phi_t$ can evolve the geodesic flow preserving the wave fronts in each point. 

The method is to solve the Reeb's vector field flow, associated to contact 1-form in the ray-optical structure to reconstruct numerically from it, the wave fronts and the ray lights propagating in an optical medium. When interfaces are present between media with different refractive index, no further adaptions are needed for the technique to reconstruct wave fronts and light rays while refracted. If this technique is used in a Minkowski spacetime geometry, 
Snell's law can be deduced and total inner reflection is obtained when the source of light is settle to refract to another medium with smaller refractive index.

\section{Wave propagation in $(2+1)$-dimensional anti-deSitter spacetime.}
\label{sec.lob}

If we consider now a $(2+1)$-dimensional anti-deSitter spacetime with spatial optical metric given by  (Ref. \citenum{LOPEZMONSALVO2020168270})
	\beq
	g=\left(\frac{n}{y}\right)^2\,\sum_{i=1}^2 \d x^i \otimes \d x^i.
	\eeq
we find the associated 1-parameter family of strict contact transformations, generated by the Reeb's flow, for which we obtained 

%	\bea
%	\label{lob.contact.trans}
%	\phi_t=&\left[x=\frac{(x\sin{\varphi}+y \cos{\varphi}-x)e^{\frac{2t}{n}}-x\sin{\varphi}-y\cos{\varphi}-x}{(\sin{\varphi}-1)e^{\frac{2t}{n}}-\sin{\varphi}-1}, y=-\frac{2y e^{\frac{t}{n}}}{(\sin{\varphi}-1)e^{\frac{2t}{n}}-\sin{\varphi}-1}\right. ,\\ \nonumber
%	&\left.\;\varphi= \arctan{\left(-\frac{(-\sin{\varphi}+1)e^{\frac{2t}{n}}-\sin{\varphi}-1}{2e^{\frac{t}{n}}\cos{\varphi}}\right)}\right]
%	\end{align}

	\bea
	\label{lob.contact.trans}
	\phi_t=&\left[x=\frac{(x\sin{\varphi}+y \cos{\varphi}-x)e^{\frac{2t}{n}}-x\sin{\varphi}-y\cos{\varphi}-x}{(\sin{\varphi}-1)e^{\frac{2t}{n}}-\sin{\varphi}-1}, \right.\\ \nonumber
	& \left. y=-\frac{2y e^{\frac{t}{n}}}{(\sin{\varphi}-1)e^{\frac{2t}{n}}-\sin{\varphi}-1}\right. ,\\ \nonumber
	&\left.\;\varphi= \arctan{\left(-\frac{(-\sin{\varphi}+1)e^{\frac{2t}{n}}-\sin{\varphi}-1}{2e^{\frac{t}{n}}\cos{\varphi}}\right)}.\right]
	\end{align}

\begin{figure}
\begin{center}
\includegraphics[width=1.05\columnwidth]{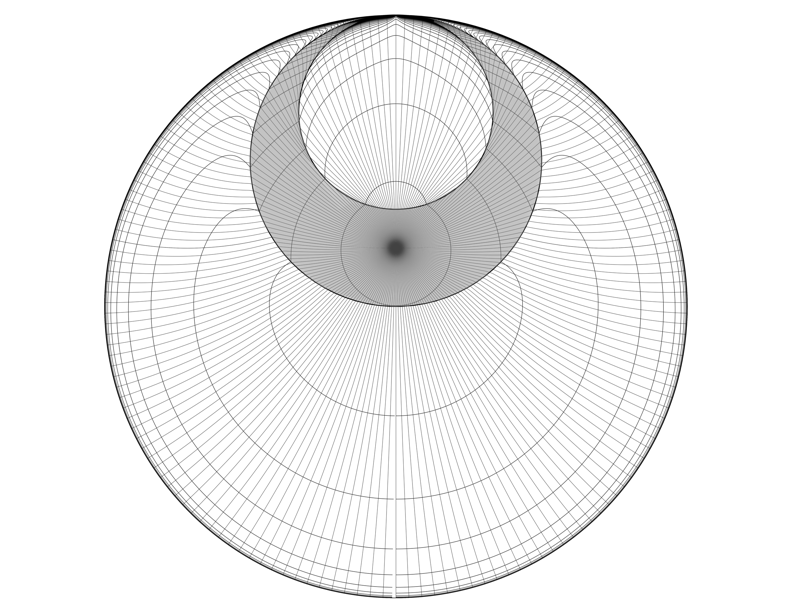}
\caption{Light rays and wavefronts emitted by a source of light refracting through a horizontal interface mapped to the Poincar\'e disc. The refractive index in the darker layer (where the light source is settled) is larger than the one in the whiter space. 
% which is in a medium
% with a refractive index larger than the outside medium, refracting through a horizontal layer with $n=1$ in the upper half plane mapped to the Poincar\'e disc. Observe that no total inner reflection is observed. As the light rays are closer to the circule boundary of the disc, some of them intersects more than one time the same wavefront, which means that light rays in different time intersects the same hypersurface of constant time. This type of aberrations are due to the geometry and can be interpreted as achronal surfaces
}
\label{hyp.tot.ref}
\end{center}
\end{figure}

With this transformation, it is possible to observe that the projection of the flows corresponds to semi-circles centered in the $x$-axis, as expected by the hyperbolic geometry associated to the anti-deSitter spacetime. For interfaces between media with different refractive index, refraction can be observed and no total inner reflection is observed when projected to the Poincar\'e disc. As it is known, the AdS spacetime is nos globally hyperbolic, so it admits closed timelike curves and achronal surfaces can be observed where light rays are closer to the infinity circle boundary of the disc. It is also observed that light rays can intersect more than one time the same wave front (figure \ref{hyp.tot.ref} Ref. \citenum{garcia2021duality}). 
 
\section{Final remarks}

The fact that geodesic flows can be written in terms of a 1-parameter family of contact transformations, allows us to reconstruct trajectories and wavefronts of light while propagating through an optical medium, just from the Reeb flow. With this technique, we recreate the wave fronts as they propagate in an optical medium with interfaces of different refractive index in a $(2+1)$-dimensional Minkowski spacetime and $(2+1)$-dimensional anti-deSitter spacetime. In the first example, we could reproduce Snell's law of refraction and the total inner reflexion phenomenon. In the second example, we observe refraction when mapped to the Poincar\'e dsic. No inner total reflection was observed when the light source is in a medium with larger refractive index. Achronal surfaces can be observed in this configuration, as light rays closer to the circle boundary intersects more than once the same wavefront.

The formalism, is robust enough to deal with interfaces without further adaptations. This technique can be used to recreate wave propagation in analogue gravitational spacetimes, such as Schwarzschild's blackholes, or to study interface between vacuum and plasma. In recent years, this type of techniques can be important in the development of metamaterials, which can mimic some spacetime properties impossible to find in nature. 

\section*{Acknowledgments}
DGP was funded by a CONACYT Scholarship with CVU 425313. 

\bibliographystyle{ws-procs961x669}
\bibliography{biblio.bib}

\end{document}